# Comments on "Efficient Band Gap Prediction for Solids" [Phys. Rev. Lett. 105, 196403 (2010)]

D. Bagayoko, G. L. Zhao, L. Franklin,
Department of Physics, Southern University and A&M College in Baton Rouge (SUBR)
Baton Rouge, Louisiana 70813, USA

E. C. Ekuma
Department of Physics and Astronomy, Louisiana State University (LSU)
Baton Rouge, Louisiana 70803, USA

An oversight of several previous local density approximation (LDA) results appears to have led to an incomplete picture of the actual capability of density functional theory (DFT), with emphasis on LDA, to describe and to predict the band gaps of semiconductors [Phys. Rev. Lett. 105, 196403 (2010)]. LDA is portrayed as failing to describe the band gap of semiconductors. In light of the content of the literature, this characterization is misleading. These comments are intended to note some of these previous results and to provide an assessment of LDA capability that is drastically different from that of failure to describe or to predict the band gaps of several semiconductors. This true capability is apparent when the required *system of equations* of DFT (or LDA) is solved self-consistently as done in the Bagayoko, Zhao, and Williams (BZW) method.

Chan and Ceder [1] reported results from an innovative approach to the calculation of electronic properties of semiconductors, including the band gap. Their Delta Self-consistent Field (ΔSCF) method, called Δ-Sol, appears to reduce by 70% the underestimation of band gaps of semiconductors as obtained in many LDA and generalized gradient approximation (GGA) calculations. As explained by the authors, the underlining physical principle of Δ-Sol consists of a consideration of the dielectric screening. The method requires one or two parameters that are to be fixed, using some reference compounds, to minimize errors in the prediction of band gaps. It also entails the use of the DFT+U scheme, with U equals to 3 eV, for the description of oxides and halides with unfilled d shells.

Chan and Ceder [1] reported markedly improved, calculated band gaps of 100 semiconductors. Specifically, the reduction of the usually woeful underestimation is 70%. Clearly, the Δ-Sol method represents a significant progress in predictive calculations of the band gaps of semiconductors. In Table II of the article [1], the authors compare their Δ-Sol calculated band gaps to corresponding, measured band gaps and to theoretical results obtained with LDA and the modified Becke and Johnson (MBJ-LDA) approach of Tran and Blaha [2].

The comparison of these band gaps in their Table II [1], for some 12 semiconductors, conveys an incomplete picture of the capability of LDA, due to the oversight of several previous LDA results obtained by solving the system of equations defining LDA [3-12]. Indeed, the LDA band gaps in this table show a very large underestimation of the corresponding, measured gaps for most of the 12 semiconductors. Throughout their article [1], the authors underscored the failure of LDA that is reportedly due to a lack of derivative discontinuity [13] for $E_{g,KS}$ and to delocalization errors. A common feature of reported LDA results that dramatically deviate from corresponding experimental ones is that the calculations were performed with single trial basis sets (STBS). Specifically, a single basis set is selected to solve the Kohn-Sham equation self-consistently. This approach does not solve the system of equations as stated by Kohn and Sham [14].

The above believed failure, understandable in light of the abundance of STBS-LDA or GGA results, is still misleading, inasmuch as it does not take into account many previous LDA band gaps that agree with experiment [3-12]. The common feature of the calculations that obtained LDA band gaps in agreement with experiment consists of solving self-consistently, in accordance with the BZW method [4, 10-12], both the Kohn-Sham equation and the equation giving the ground state charge density in terms of the wave functions of the occupied states. Once a potential is selected, the system of equations describing DFT reduces to the above two equations. For clarity, we shall label our results with LDA-BZW, to distinguish them from the very abundant outcomes from STBS-LDA calculations.

The results shown in Table I below, for Δ-Sol [1] and MBJ-LDA [2] calculations, are the ones in Table II of Chan and Ceder [1]. Also shown in Table I are some corresponding LDA-BZW results, including two very recent ones, for ZnO [15] and CdS [16], that have just been submitted for publication. The content of the table clearly indicates that Δ-Sol and MBJ-LDA calculations do not outperform LDA-BZW ones for the description of the band gaps of semiconductors. The case of wurtzite GaN is particularly illustrative. While Δ-Sol and MBJ-LDA calculations reported 3.9 and 2.8 eV for the band gap, respectively, LDA-BZW calculations found 3.2 eV that is much closer to the experimental value of 3.4 eV.

Table II shows several other LDA-BZW results, in the literature, that agree with experiment. In particular, our predictions for the bulk modulus and the band gap for cubic silicon nitrate [6], published before any measurements were known to us, were confirmed by experiment [17, 18]. Similarly, our predicted lattice constant and band gap for cubic indium nitride [7] agree very well with corresponding measured values [19]. Several of the articles containing these results also reported LDA-BZW calculated electron effective masses in agreement with experiment. As noted elsewhere [3-6, 12], this agreement is a measure of the correctness of the shape of the conduction band near its minimum. Our recent works on ZnO and CdS led to peaks in the density of states that agree with measurements, not only for the valence bands, but also for the low laying conduction bands. This agreement is not surprising, in light of the work of Bagayoko et al. [3] and of Jin et al. [9] who reported calculated optical properties in agreement with experiment up to energies of 5.5 to 6 eV. This latter agreement denotes not only the correct description of the valence bands, but also that of low laying conduction bands – including their separations from the valence bands.

Our BZW calculations employed the linear combination of atomic orbitals (LCAO) formalism whose implementation followed the Bagayoko, Zhao, and Williams (BZW) method. This method, starting with the minimum basis set, systematically increases the size of the basis set to perform successive, self-consistent calculations of the band structure. The comparison of the occupied energies of adjacent calculations ultimately leads to the identification of the *optimal basis set*, the one beyond which the occupied energies no longer change (nor do the charge density and the potential). This optimal basis set is the smallest one that gives the minima of all the occupied energies. These minima are the same as obtained with larger basis sets resulting from augmenting the optimal basis set. The optimal basis set is complete for the description of the ground state and is not over-complete for it, like larger basis sets. Essentially, our totally ab-initio calculations owe their only distinction to solving self-consistently both the Kohn-Sham equation and the equation giving the ground state charge density in terms of the wave functions of the occupied states as thoroughly described elsewhere [4, 10-12]. We utilized Gaussian functions to describe the radial parts of the orbitals. Most of the calculations employed the Ceperley and Alder [20] local density potential as parameterized by Vosko, Wilk and Nusair [21]. For wurtzite InN [8] and zinc blende AlAs [22], we also utilized the generalized gradient approximation potential of Perdew et al.[23-24].

The contents of some of our articles [10-12] established the fact that the sources of the reported failure of STBS-LDA calculations are far from being well understood. In particular, it has not been proven that the derivative discontinuity of the exchange correlation energy is positive in real semiconductors; Sham and Schlüter [13] explicitly stated that their work does not demonstrate whether or not it is zero in insulators. Further, the original derivations of DFT [25] and of LDA [14] have meanings only for the ground state. Hence, STBS-LDA calculations that (a) do not verifiably guarantee the completeness of the basis set for the description of the ground state and that (b) deliberately and explicitly attempt to get "converged" excited state energies out of a totally ground state system of equations do not often provide a full picture of the capability of DFT or LDA. Item (b) above is equivalent to not avoiding over-complete basis sets for the description of the ground state. As such, the outcomes of STBS-DFT or -LDA calculations do not represent the true capability of this ground state theory whose correct utilization, for electronic structure calculations, requires the self-consistent solution of the pertinent *system of equations* as stated by Kohn and Sham [14] and by Kohn [26].

In summary, it appears that the oversight of the available LDA-BZW results in Tables I and II led to an incomplete picture of the capability of DFT and LDA to describe and to predict electronic and related properties of semiconductors, including band gaps. While Δ-Sol, MBJ-LDA, and GW approaches lead to markedly better band gaps for semiconductors, particularly as compared to results from STBS-LDA calculations, we are not aware of any scheme to date that outperforms LDA-BZW method. In particular, our method has led not just to correct band gaps, but also to densities of states, optical properties, and electron effective masses in agreement with experiment. The LDA-BZW predictions noted above have been validated by measurements. The above data and related discussions clearly point to the need to revisit previously believed limitations of DFT and of LDA, based on results of STBS calculations, and to realize that solutions of the DFT (or LDA) systems of equation have much more physical meaning than intimated or suggested in the literature.

**Acknowledgments**

This work was funded in part by the Louisiana Optical Network Initiative (LONI, Award No. 1110005204), the National Science Foundation and the Louisiana Board of Regents (Award Nos. EPS-1003897 and NSF(2010-15)-RII-SUBR), and by Ebonyi State, Federal Republic of Nigeria (Award No. EBSG/SSB/FSA/040/VOL. VIII/039).

**Table I.** A comparison of results from Δ-Sol$_{LDA}$, MBJ-LDA, and LDA-BZW calculations with experimental measurements. Except for the low temperature band gap of 3.51 eV for wurtzite ZnO, the data in columns 1 through 5 are as in Table II of Reference 1. The band gaps ($E_g$) are in electron volts (eV).

| Compound | $E_g$ (Experiment) | $E_g$ KS$_{LDA}$ | $E_g$ MBJ-KS$_{LDA}$ | $E_g$ Δ-Sol$_{LDA}$ | $E_g$ LDA-BZW |
|---|---|---|---|---|---|
| C | 5.5 | 4.1 | 4.9 | 5.3 | 5.05[a] |
| Si | 1.1 | 0.5 | 1.2 | 1.0 | 1.02[a] |
| Ge | 0.7 | 0.0 | 0.9 | 0.9 | 0.6[b] |
| SiC - 3C SiC | 2.2 | 1.4 | 2.3 | 2.4 | 2.24[c] |
| BN(cubic) | 6.2 | 4.4 | 5.9 | 5.8 | |
| GaN | 3.4 | 1.6 | 2.8 | 3.9 | 3.2[a] |
| GaAs | 1.4 | 0.3 | 1.6 | 1.5 | |
| AlP | 2.5 | 1.5 | 2.3 | 2.1 | |
| ZnS | 3.7 | 1.8 | 3.7 | 3.6 | |
| CdS | 2.5 | 0.9 | 2.7 | 3.0 | 2.47[d] |
| AlN | 6.1 | 4.2 | 5.6 | 5.3 | |
| ZnO | 3.3 & 3.51[*] | 0.8 | 2.7 | 3.5 | 3.47[e] |

[*] Very low temperature band gap (i.e., T= 0), as cited in Reference [15].

[a]Reference [4]    [b]Unpublished    [c]Reference [5]    [d]Reference [16]

[e]Reference [15]

**Table II.** Some other LDA-BZW results available in the literature and that are not surpassed, to our knowledge, by STBS-LDA, modified LDA (i.e., LDA+C, LDA+U, etc.), GW, or other calculations to date, as far as the correct description of the noted materials is concerned. The band gaps ($E_g$) are in electron volts (eV).

| Compound | $E_g$ (Experiment) – Unless otherwise indicated | $E_g$ (LDA-BZW) - Unless otherwise indicated |
|---|---|---|
| BaTiO$_3$ | 2.8 - 3.0 | 2.6[a] |
| 4H-SiC | 3.2 & 3.3 | 3.11[b] |
| Si$_3$N$_4$ (Cubic)[*] | Bulk Modulus: 317 ± 11 GPa[c] | Bulk Modulus: 330 GPa[d] |
| | 3.6 - 3.7[e] | 3.68[d] |
| InN (wurtzite) | 0.7-1.0[f] | 0.88[g] |
| | | 0.81 (GGA-BZW)[g] |
| InN (Cubic)[**] | Lattice Constant: 5.01 ± 0.01 Å[h] | Lattice Constant: 5.017 Å[i] |
| | 0.61[h] | 0.65[i] |
| AlAs (zinc blende- zb) | Room Temperature gaps: 2.15- 2.16 | 2.15 (from GGA-BZW)[j] |
| SWCNT (10,0) | 0.90-0.96 | 0.95[k] |

[*] The 2001 prediction of a bulk modulus was confirmed by experiment in 2002 and that for the band gap of 3.68 eV was confirmed in 2003.

[**] The 2004 predicted lattice constant and band gap were confirmed by experiment in 2006.

[a]Reference [3]      [b]Reference [5]      [c]Reference [17]      [d]Reference [6]

[e]Reference [18]      [f]As explained in Reference [8], with the Burstein-Moss Shift

[g]Reference [8]      [h]Reference [19]      [i]Reference [7]      [j]Reference [22]

[k]Reference [27]